\documentclass[aps,prd,twocolumn,superscriptaddress,amssymb,eqsecnum,showpacs,showkeyes,secnumarabic,graphics,floatfix,nofootinbib,tightenlines,longbibliography]{revtex4-1}

\usepackage{graphicx}
\usepackage{epsfig}
\usepackage{bm}
\usepackage{dcolumn}
\usepackage{amsmath}
\usepackage{hhline}
\usepackage{multirow}
\usepackage{longtable}
\usepackage[utf8]{inputenc}
\usepackage[breaklinks=true,colorlinks=true,
linkcolor=blue,urlcolor=blue,citecolor=cyan,
bookmarks=true,bookmarksopenlevel=2]{hyperref}

\def \beq  {\begin{equation}}
\def \eeq  {\end{equation}}
\def \ber  {\begin{eqnarray}}
\def \eer  {\end{eqnarray}}

\begin{document}
\newcommand{\newc}{\newcommand}

\newc{\be}{\begin{equation}}
\newc{\ee}{\end{equation}}
\newc{\ba}{\begin{eqnarray}}
\newc{\ea}{\end{eqnarray}}
\newc{\bea}{\begin{eqnarray*}}
\newc{\eea}{\end{eqnarray*}}
\newc{\D}{\partial}
\newc{\ie}{{\it i.e.} }
\newc{\eg}{{\it e.g.} }
\newc{\etc}{{\it etc.} }
\newc{\etal}{{\it et al.}}
\newc{\lcdm}{$\Lambda$CDM }
\newcommand{\nn}{\nonumber}
\newc{\ra}{\Rightarrow}
\newc{\lsim}{\buildrel{<}\over{\sim}}
\newc{\gsim}{\buildrel{>}\over{\sim}}

\title{Cosmological Horizons, Uncertainty Principle and Maximum Length Quantum Mechanics}

\author{ L. Perivolaropoulos} \email{leandros@uoi.gr}
\affiliation{Department of Physics, University of Patras, 26500 Patras, Greece \\(on leave from the Department of Physics, University of Ioannina, 45110 Ioannina, Greece) }

\date {\today}

\begin{abstract}
The cosmological particle horizon is the maximum measurable length in the Universe. The existence of such a maximum observable length scale implies a modification of the quantum uncertainty principle. Thus due to non-locality of quantum mechanics, the global properties of the Universe could produce a signature on the behaviour of local  quantum systems. A Generalized Uncertainty Principle (GUP) that is consistent with the existence of such a maximum observable length scale $l_{max}$ is $\Delta x \Delta p \geq \frac{\hbar}{2}\;\frac{1}{1-\alpha \Delta x^2}$ where $\alpha = l_{max}^{-2}\simeq (H_0/c)^2$ ($H_0$ is the Hubble parameter and $c$ is the speed of light). In addition to the existence of a maximum measurable length $l_{max}=\frac{1}{\sqrt \alpha}$, this form of GUP implies  also the  existence of a minimum measurable momentum $p_{min}=\frac{3 \sqrt{3}}{4}\hbar \sqrt{\alpha}$. Using appropriate representation of the position and momentum quantum operators we show that the spectrum of the one dimensional harmonic oscillator   becomes $\bar{\mathcal{E}}_n=2n+1+\lambda_n \bar{\alpha}$ where $\bar{\mathcal{E}}_n\equiv 2E_n/\hbar \omega$ is the dimensionless properly normalized $n^{th}$ energy level, $\bar{\alpha}$ is a dimensionless parameter with $\bar{\alpha}\equiv \alpha \hbar/m \omega$ and $\lambda_n\sim n^2$ for $n\gg 1$ (we show the full form of $\lambda_n$ in the text). For a typical vibrating diatomic molecule and $l_{max}=c/H_0$ we find $\bar{\alpha}\sim 10^{-77}$ and therefore for such a system, this effect is beyond reach of current experiments. However, this effect could be more important in the early universe and could produce signatures in the  primordial perturbation spectrum induced by quantum fluctuations of the inflaton field. 

\end{abstract}
\maketitle

\section{Introduction}
\label{sec:Introduction}

Quantum Theory (QT) has been tested to a great extend in the context of microphysical systems and has been shown to be consistent with all current experiments. It is a self-consistent and well established theory. Despite of the significant success of QT there are two issues that appear to challenge the theory:
\begin{enumerate}
\item
QT appears to be incompatible with General Relativity (GR) due to non-renormalizable divergences that appear when GR is quantized. This incompatibility implies that at least one of the two theories (GR or QT) needs to be modified.
\item
There is no clear and unique interpretation of QT. Even though QT has withstood  rigorous and thorough experimental testing, the outcomes of these experiments are open to different interpretations of physical reality.
\end{enumerate}
It is therefore clear that a possible generalization of QT is a viable and interesting prospect. Such a generalization would most likely affect the cornerstone of QT that effectively defines it: the Heisenberg Uncertainty Principle\cite{aHeisenberg:1927zz,Robertson:1929zz} (HUP) converting it to a Generalized Uncertainty Principle (GUP)\cite{Maggiore:1993rv,Tawfik:2015rva}. 

A well motivated form of GUP is based on the assumption of the existence of a fundamental ultraviolet cutoff or equivalently a  Minimum Measurable Length. This assumption has been suggested in quantum gravity \cite{Garay:1994en,AmelinoCamelia:2008qg,Hossenfelder:2012jw-minx-qg,Adler:1999bu}, quantum geometry \cite{Capozziello:1999wx} as well as in string theory\cite{Veneziano:1986zf,Amati:1987wq,Gross:1987kza,Konishi:1989wk,Kato:1990bd}. It is based on
the expectation that high energies used in the resolution of  small scales will lead to significant disturbances of spacetime structure by their gravitational effects\citep{Plato:2016azz-thought-exp-smallest-scale-gravity-effects}. Such a disturbance which may take the form of a black hole, could prohibit the probe of scales smaller than a cutoff which is expected to be of the order of the Planck scale. Thus, the coexistence of QT with GR naturally leads to the requirement of a modification of both QT and GR, the introduction of a fundamental ultraviolet cutoff and thus a GUP consistent with both a Minimum Measurable Length and a Maximum Measurable Momentum (ultraviolet cutoff). These effects are integrated in the GUP as {\it minimum position}\cite{Hossenfelder:2012jw-minimal-length-review,Kempf:1994su-basic,Hossenfelder:2012jw-minx-qg,Das:2008kaa-3dGUP-exp-bounds,Kempf:1996fz-3dgup-harm-osc,Tawfik:2014zca-gup-review,Kempf:1994qp} and {\it maximum momentum}\cite{Ali:2011fa,Das:2008kaa-3dGUP-exp-bounds,Cortes:2004qngup-in-dsr,Ali:2009zq-GUP-discrete-space,Nozari:2012gd-general-gup-tunelling-bh} uncertainty.

This type of GUP has been extensively studied since the pioneering work of Ref. \cite{Kempf:1994su-basic}  that introduced it in the form
\be
\Delta x \Delta p \geq \frac{\hbar}{2} (1+\beta \Delta p^2)
\label{gup1}
\ee
where $\beta$ is  the GUP parameter defined as $\beta =\beta_0 /M_{pl} c^2=\beta_0 l_{pl}^2/\hbar^2$,  $M_{pl}c^2=10^{19}GeV$, $l_{pl}=10^{-35}m$ is the
4-dimensional fundamental Planck scale and $\beta_0$ is a dimensionless parameter expected to be of order unity. Estimates of the values of $\beta_0$ may be obtained by using leading quantum corrections to the Newtonian potential \cite{Scardigli:2016pjs-gup-modif-newt-potl}. At energies much lower than the Planck energy the $\beta$ correction of the GUP becomes negligible and the HUP is recovered.

The minimum allowed position uncertainty obtained from the GUP (\ref{gup1}) is obtained for a finite value of $\Delta p= \frac{1}{\sqrt \beta}$ and corresponds to a minimum position uncertainty different from zero ($\Delta x_{min}=2 \hbar \sqrt{\beta}$)\cite{Pedram:2011xj-schrod-eq-min-length-max-mom}. The GUP (\ref{gup1}) is obtained from a generalized Heisenberg algebra\cite{Kempf:1994su-basic} as discussed in the next section.

\begin{figure}[!t]
\centering
\vspace{0cm}\rotatebox{0}{\vspace{0cm}\hspace{0cm}
\resizebox{0.49\textwidth}{!}{\includegraphics{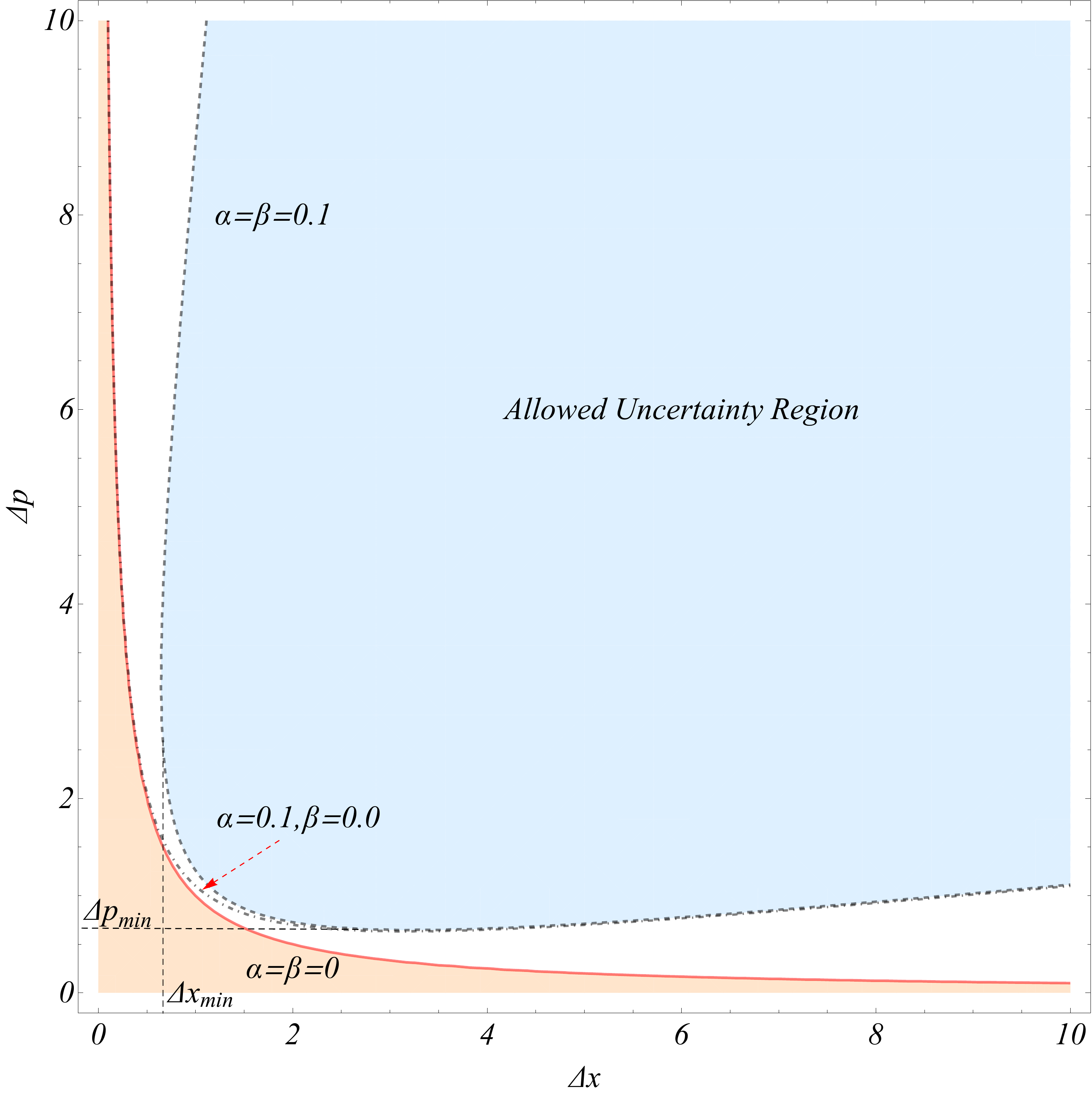}}}
\caption{The deformation of the HUP in the presence of the parameters $\alpha$ and $\beta$. The Figure shows the allowed uncertainty region assuming the GUP of eq. (\ref{gup2}) with a minimum position and a minimum momentum uncertainty.}
\label{figgupminpminx}
\end{figure}

A natural generalization of (\ref{gup1}) corresponds to the existence of minimum position and minimum momentum uncertainty\cite{CostaFilho:2016wvf-min-momentum}. This is obtained by a GUP of the form\cite{Kempf:1994su-basic,Bojowald:2011jd-GUP-discrete-space}
\be
\Delta x \Delta p \geq \frac{\hbar}{2} (1+\alpha \Delta x^2+\beta \Delta p^2)
\label{gup2}
\ee
The fact that this form of GUP predicts the existence of both a minimum position and a minimum momentum uncertainty is illustrated in Fig. \ref{figgupminpminx}  where we show that the deformation of the HUP due to the introduction of the parameters $\alpha$ and $\beta$ leading to minimum uncertainties for both momentum and position. The uncertainties $\Delta x$, $\Delta p$ and the parameters $\alpha$, $\beta$ in Fig. \ref{figgupminpminx} have been rescaled to dimensionless form by appropriate microphysical scales $l_{mp}$ and $p_{mp}\equiv \frac{\hbar}{2 l_{mp}}$ which depend on the micropysical system under consideration. For example for a harmonic oscillator we have $l_{mp}=\sqrt{\hbar/m\omega}$.  This rescaling may be expressed as $\frac{\Delta x}{l_{mp}} \rightarrow \Delta x$, $\frac{\Delta p}{p_{mp}} \rightarrow \Delta p$, $\alpha l_{mp}^2 \rightarrow \alpha$ and $\beta p_{mp}^2 \rightarrow \beta$.

\begin{figure*}[ht]
\centering
\begin{center}
$\begin{array}{@{\hspace{-0.10in}}c@{\hspace{0.0in}}c}
\multicolumn{1}{l}{\mbox{}} &
\multicolumn{1}{l}{\mbox{}} \\ [-0.2in]
\epsfxsize=3.3in
\epsffile{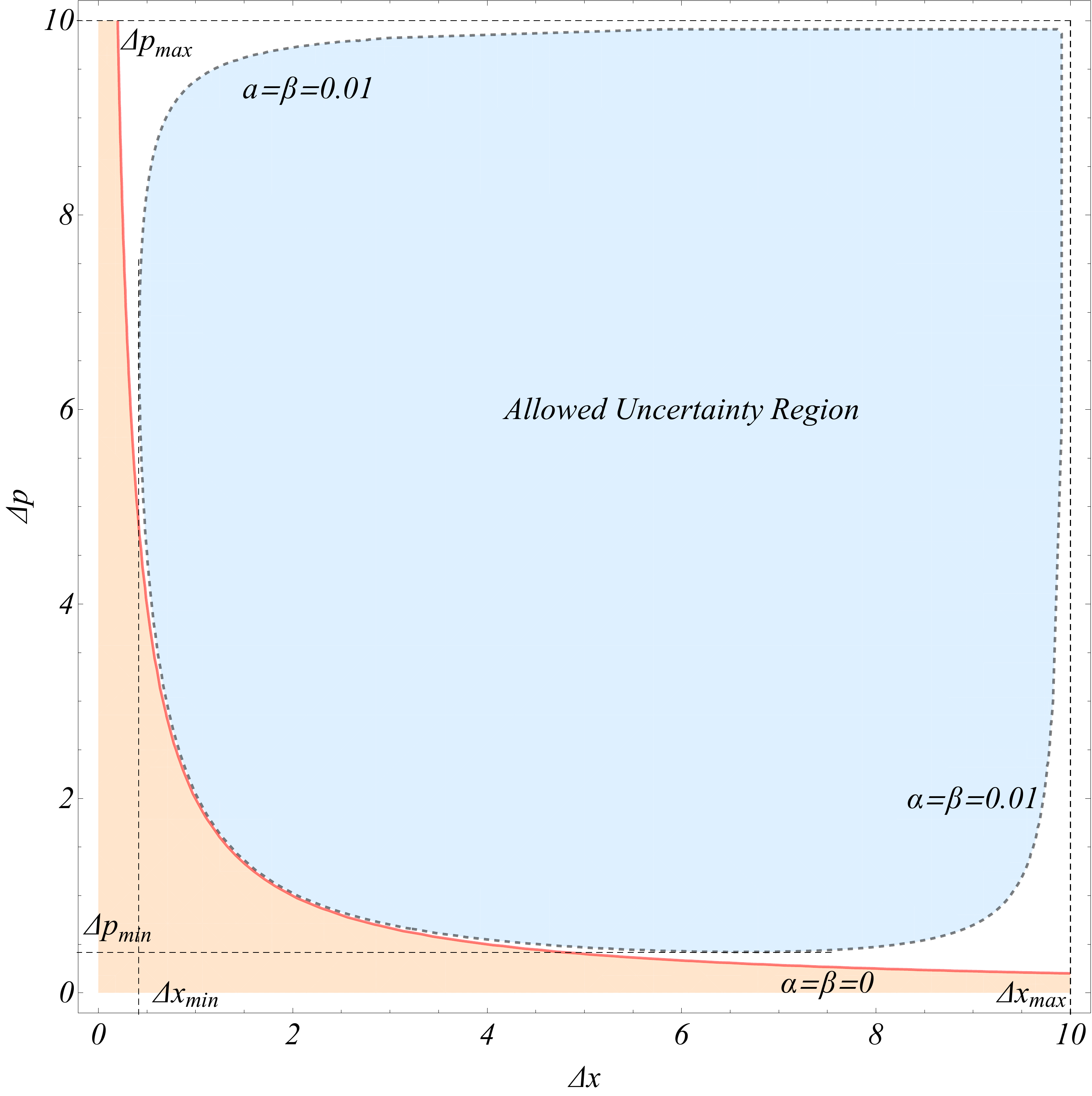} &
\epsfxsize=3.3in
\epsffile{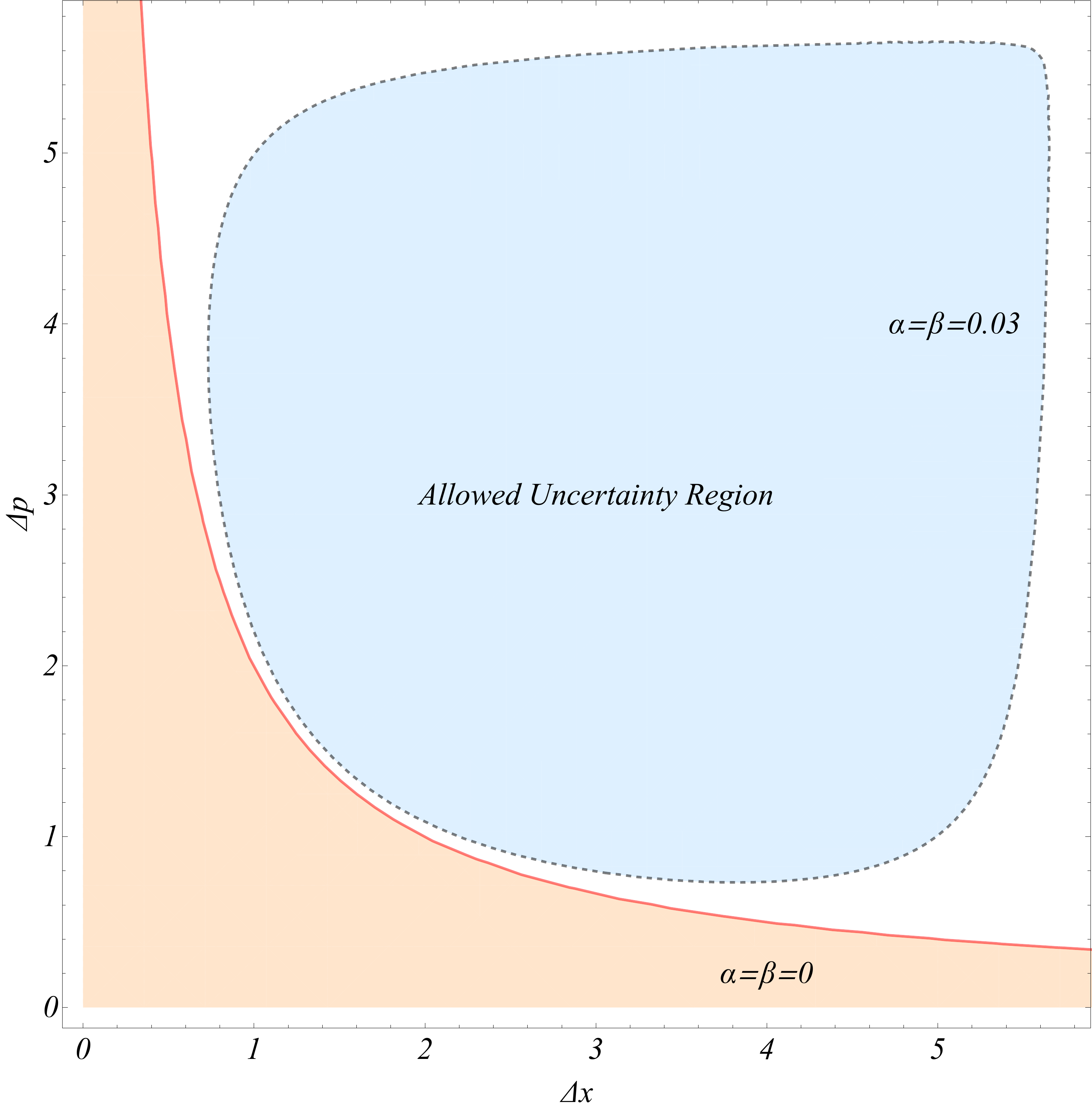} \\
\end{array}$
\end{center}
\vspace{0.0cm}
\caption{The deformation of the HUP in the presence of the parameters $\alpha$ and $\beta$ assuming the GUP of eq. (\ref{gupexpmaxmommaxpos}) with both minima and maxima in position and momentum uncertainties. The uncertainties $\Delta x$, $\Delta p$ and the parameters $\alpha$, $\beta$ have been rescaled to dimensionless form by appropriate microphysical scales $l_{mp}$ and $p_{mp}\equiv \frac{\hbar}{l_{mp}}$ ($\frac{\Delta x}{l_{mp}} \rightarrow \Delta x$, $\frac{\Delta p}{p_{mp}} \rightarrow \Delta p$, $\alpha l_{mp}^2 \rightarrow \alpha$ and $\beta p_{mp}^2 \rightarrow \beta$). The left panel shows the allowed uncertainty region for $\alpha=\beta=0.01$ while for the right panel we have $\alpha=\beta=0.03$ leading to a smaller allowed uncertainty region. }
\label{figgupminmaxxp}
\end{figure*}

Any form of GUP implies the existence of a deformed Heisenberg algebra. For example a GUP of the form (\ref{gup2}) implies a phase space commutator in one dimension of the form:
\be
[x,p]=\frac{\hbar}{2}(1+\alpha x^2 + \beta p^2)
\label{comrel1}
\ee
As discussed in the next section, in higher dimensions this commutation relation becomes more complicated\cite{Kempf:1994su-basic} if we want to keep a commutative geometry ($[x_i,x_j]=[p_i,p_j]=0$).

An alternative form of GUP is motivated by the fact that if a fundamental minimal length indeed exists in Nature it should also have the property of being invariant with respect to Lorentz transformations. This requires also a deformation of the Lorentz group and a nonlinear modification of Lorentz transformations. This corresponds to a modification of Special Relativity to a theory known as {\it Doubly Special Relativity} (DSR) \cite{Magueijo:2001cr,Magueijo:2002am-dsr,Cortes:2004qngup-in-dsr,Magueijo:2004vv}. In this class of theories, Lorentz transformations are generalized to a form
\ba
E' &=& f(E,p,l_{min},v) \\
p' &=& g(E,p,l_{min},v) 
\label{dsr}
\ea
where $(E,p)$ are energy and momentum, $l_{min}$ is the invariant minimal length scale expected to be of the order of the Planck scale and $v$ is the velocity of the transformation. The functions $f$ and $g$ are selected so that the length scale $l_{min}$ remains invariant with respect to the new modified Lorentz transformations and are severely constrained by experiments/observations\cite{Hees:2016lyw,Liberati:2013xla}. It can be shown \cite{Cortes:2004qngup-in-dsr} that in this class of models there is a natural ultraviolet (UV) cutoff of momentum while the commutation relation in one dimension gets generalized by the addition of a linear term to the form\cite{Ali:2011fa,Das:2008kaa-3dGUP-exp-bounds,Cortes:2004qngup-in-dsr,Ali:2009zq-GUP-discrete-space,Nozari:2012gd-general-gup-tunelling-bh}
\be
[x,p]=i \hbar (1-\beta_1 p + 2\beta_1^2 p^2)
\label{comrel2}
\ee
while the uncertainty principle takes the form 
\be
\Delta x \Delta p \geq \frac{\hbar}{2}  (1-2 \beta_1 < p >+ 4\beta_1^2 <p^2>)
\label{gup2a}
\ee
where the subscript $_1$ is used to differentiate $\beta_1$ from the parameter $\beta$ which has different dimensions.
In this form of GUP there is no explicit UV cutoff in the momentum uncertainty even though there is an implicit such cutoff through arguments related to  DSR \cite{Cortes:2004qngup-in-dsr}. 

An explicit UV cutoff can be obtained through the GUP\citep{Pedram:2011gw-true-max-mom,Pedram:2011xj-schrod-eq-min-length-max-mom}
\be
\Delta x \Delta p \geq \frac{\hbar}{2}  \frac{1}{1- \beta \Delta p^2}
\label{gupexpmaxmom}
\ee
It originates from a commutation relation of the form
\be
[x,p]=i \hbar \frac{1}{1-\beta p^2}
\label{comrelexpmaxmom}
\ee
The GUP of eq. (\ref{gupexpmaxmom}) can be further generalized to include explicit maxima and minima in both position and momentum uncertainties. We thus obtain a GUP of the form
\be
\Delta x \Delta p \geq \frac{\hbar}{2}  \frac{1}{1- \beta \Delta p^2} \frac{1}{1- \alpha \Delta x^2}
\label{gupexpmaxmommaxpos}
\ee
The allowed region of uncertainties of this very general deformed GUP is shown in Fig. \ref{figgupminmaxxp} (light blue region) where $\Delta x$ and $\Delta p$ have been rescaled to dimensionless form by appropriate microphysical scales $l_{mp}$ and $p_{mp}$ defined above.

The presence of an infrared cutoff GUP (explicit presence of a maximum position uncertainty) as implemented in eq. (\ref{gupexpmaxmommaxpos}) has not been considered previously in the literature to our knowledge. However, there is a well defined motivation for such a cutoff in the context of either cosmological particle horizons\cite{Faraoni:2011hf-cosmic-horizons,Davis:2003ze-horizons} or non-trivial cosmic topology\citep{Luminet:2016bqv-cosmic-topology-constr} which provide a maximum measurable length scale in the Universe. In particular, the particle horizon corresponds to the length scale of the boundary between the observable and the unobservable regions of the universe. This scale at any time defines the size of the observable universe. The physical distance to this maximum observable scale at the cosmic time $t$ is given by
\be
l_{max}(t)=a(t)\int_0^t \frac{c\; dt}{a(t)}
\label{parthorscale}
\ee
where $a(t)$ is the cosmic scale factor.
For the best fit \lcdm cosmic background at the present time $t_0$ we have
\be
l_{max}(t_0)\simeq 14 Gpc \simeq 10^{26} m
\label{parthort0}
\ee
In the context of the presence of such an infrared cutoff the following questions arise:
\begin{itemize}
\item
What are the possible forms of GUP that include an infrared cutoff in the form of a maximum measurable length and therefore a maximum position uncertainty?
\item
What are the experimental predictions of the corresponding generalized quantum theories for simple quantum systems?
\item
What are the theoretical/observational predictions of the corresponding generalized quantum theories for black hole thermodynamics?
\item
Are there cosmological signatures predicted by such GUP?
\end{itemize}
The discussion of some of these questions and the proposal of possible answers is the focus of the present analysis. 

The structure of this paper is the following: In the next section we review the basic forms of GUP that have been analysed in the literature in one and three dimensions. We review the construction of operator representation for each form of GUP and the analysis of simple quantum systems. In section III.1 we focus on the particular form of GUP that is consistent  with a maximum measurable length scale and thus a maximum position uncertainty (Maximum Length Quantum Mechanics). We show that this form of GUP naturally also implies the existence of a minimum momentum uncertainty and derive the position-momentum operator representation of this theory in terms of the usual position-momentum operators. In section III.2 we solve the harmonic oscillator problem in the new theory and derive the spectrum as a function of the maximum observable length scale. In section III.3 we briefly discuss the expected time dependence of the maximum position uncertainty. Finally in section IV we conclude, summarize and discuss future prospects of this work.

\section{Review of Minimum Length Quantum Mechanics}
\label{sec:Section 2}

\subsection{One Space Dimension}

It is straightforward to derive the GUP eq. (\ref{gup1}) using the generalized commutation relation
\be
[x,p]=i\hbar(1 + \beta p^2)
\label{comrel1}
\ee

Using the general uncertainty principle for any pair of non-commuting observables $A$, $B$
\be
\Delta A \; \Delta B \geq \frac{1}{2} \vert <[{\hat A},{\hat B}]>\vert 
\label{obsuncert}
\ee
where $\Delta A\equiv \sqrt{<({\hat A} - <{\hat A}>)^2>}$ (similar for $B$) and ${\hat A}$, ${\hat B}$ are the operator representations of the observables $A$ and $B$. Using eq. (\ref{comrel1}) in eq. (\ref{obsuncert}) we find
\be
\Delta x \Delta p \geq \frac{\hbar}{2}  (1+ \beta \Delta p^2 +\beta <p>^2)
\label{gupxmin1}
\ee
which leads to the GUP of eq. (\ref{gup1}). 
As discussed in the Introduction, this equation may be written in the dimensionless form
\be
\Delta {\bar x} \Delta {\bar p} \geq  (1+{\bar \beta} \Delta {\bar p}^2)
\label{gup1dimless}
\ee
where ${\bar x}\equiv \frac{x}{l_{mp}}$, ${\bar p}\equiv \frac{p}{p_{mp}}$ and ${\bar \beta}\equiv \beta p_{mp}^2$. In what follows we omit the bar but we use the dimensionless form of the GUP.

The equation saturating the GUP inequality (\ref{gup1dimless}) may be written as
\be
\Delta x = \beta \Delta p + \frac{1}{\Delta p}
\label{gup1a}
\ee
It is easy to see that $\Delta x$ is minimized for $\Delta p =\frac{1}{\sqrt{\beta}}$  and the corresponding minimum position uncertainty is 
\be 
\Delta x_{min}= 2\sqrt{\beta}
\label{deltaxmin}
\ee

The operator representation that leads to the commutation relation (\ref{comrel1}) is not uniquely obtained. The position and momentum operators that obey (\ref{comrel1}) may be defined in terms of operators $x_0$, $p_0$ that obey the usual commutation relation $[x_0,p_0]=i \hbar$ as
\ba 
x&=& x_0 
\label{repr1a}\\
p&=& p_0(1+\frac{\beta}{3} p_0^2)
\label{repr1}
\ea
An alternative representation is
\ba 
x&=&(1+\beta p_0^2) x_0 
\label{repr2a}\\
p&=& p_0 
\label{repr2}
\ea
It is easy to show that both representations (\ref{repr1a})-(\ref{repr1}) and (\ref{repr2a})-(\ref{repr2}) satisfy the generalized commutation relation (\ref{comrel1}) to $O(\beta)$.. Both operator representations may be used to construct and solve a generalized Schrodinger equation for simple quantum mechanical systems\cite{Das:2008kaa-3dGUP-exp-bounds,Das:2009hs-3dGUP-exp-bounds,Fityo:2005xaa-gup-qmprobs,Quesne:2006ue-gen-gup-qmprobs,Samar:2015bqx-gup-minx-qmproblemssolved,Nozari:2005ex,Bawaj:2014cda} in one space dimension leading to generalized spectra that are consistent with the existence of a fundamental minimum lengthscale \cite{Kempf:1994su-basic}. They may also be used to derive thermodynamics properties of gravity and black holes \cite{Nozari:2011gj,Zhang:2015gda,Scardigli:2016pjs-gup-modif-newt-potl,Tawfik:2015kga,Casadio:2014pia,Ali:2013ii,Ali:2013ma,Ali:2015zua}

The operator representation (\ref{repr1a})-(\ref{repr1}) is more suitable for perturbative analysis of quantum systems while in the representation (\ref{repr2a})-(\ref{repr2}) the Hamiltonian eigenvalue problems may usually be expressed as a relatively simpler second order ODE in momentum space which may lead to exact generalized solutions \cite{Kempf:1994su-basic}.

The more general GUP (\ref{gup2a}) inspired from DSR may also be written after proper rescaling in the form
\be 
{\Delta x} \geq \frac{1}{\Delta p}-\beta_1 + \beta_1^2 \Delta p
\label{gup2a}
\ee
which is easily shown to lead to a minimum position uncertainty $\Delta x_{min}=\beta_1$ which is obtained when $\Delta p=\frac{1}{\beta_1}$.

\subsection{Three Dimensions}

A naive generalization of the commutation relation (\ref{comrel1}) to three dimensions would correspond to a commutation relation of the form
\be
[x_i,p_j]=\frac{\hbar}{2} \delta_{ij} (1 + \beta p^2)
\label{comrel13dwr}
\ee
In the context of the Jacobi identity however, this generalization would lead to non-commutative geometries ($[x_i,x_j]\neq 0$). In order to restore commutativity of spatial coordinates the above commutation relation should be generalized to\cite{Kempf:1994su-basic,Tawfik:2014zca-gup-review}
\be
[x_i,p_j]=\frac{\hbar}{2}  \left[\delta_{ij} (1 + \beta p^2)+\beta' p_i p_j \right]
\label{comrel13dcor}
\ee
where the parameter $\beta'$ is connected to the parameter $\beta$ by demanding commutativity of position vector components 
\be 
[x_i,x_j]= 0
\label{poscom}
\ee
and of  momentum vector components 
\be 
[p_i,p_j]= 0
\label{momcom}
\ee
in the context of the Jacobi identity
\be 
[[x_i,x_j],p_k]+[[x_j,p_k],x_i]+[[p_k,x_i],x_j]=0
\label{jacobi}
\ee
It is straightforward to show that to lowest order in $\beta$ and $\beta'$ equations (\ref{poscom}) and (\ref{jacobi}) imply that 
\be 
\beta'=2\beta
\label{betaprime}
\ee
Equation (\ref{betaprime}) may also be obtained using (\ref{momcom}) and the Jacobi identity of the form 
\be 
[[p_i,p_j],x_k]+[[p_j,x_k],p_i]+[[x_k,p_i],p_j]=0
\label{jacobi1}
\ee
For a general $\beta'$ the commutators of the position vector components may be shown to take the form\cite{Tawfik:2014zca-gup-review}
\be 
[x_i,x_j]=i\hbar \frac{(2\beta-\beta')+(2\beta+\beta')\beta p^2}{1+\beta p^2}(p_ix_j - p_j x_i)
\label{xixjcom}
\ee
which goes to 0 as expected for $\beta'=2\beta$ to first order in $\beta$ and $\beta'$.

The representation of position and momentum operators that is consistent with (\ref{comrel13dcor}) with (\ref{betaprime}) is of the form
\ba 
x_i&=& x_{0i}
\label{repr3dx}  \\
p_i&=& p_{0i}(1+\beta p_0^2)
\label{repr3dp}
\ea
where $x_{0i}$ and $p_{0i}$ satisfy the HUP commutations relations.

The representation (\ref{repr3dx}), (\ref{repr3dp}) may be used\cite{Benczik:2005bh-gup-hydrogen,Brau:1999uv-hydrogen,Kempf:1996fz-3dgup-harm-osc,Bufalo:2016wpm,Gao:2016fmk,Vinas:2016jmp,Castellanos:2015bia} to derive the spectra of simple quantum mechanical systems whose dynamics is determined for example by  central potentials. In such systems the Hamiltonian is of the form\citep{Brau:1999uv-hydrogen}
\be
H=\frac{p^2}{2m}+V(r)=\frac{p_0^2(1+\beta p_0^2)^2}{2m}+V(r)
\label{hamilt1}
\ee
The energy eigenvalue problem
\be 
H |\Psi_k> = E_k |\Psi_k>
\label{eigval1}
\ee
may be solved perturbatively setting $E_k = E_k^0 + \Delta E_k$ with unperturbed ($\beta =0$) states $|\Psi_k^0>$. The energy eigenvalue shifts $\Delta E_k$ are the eigenvalues of the matrix
\be 
\frac{\beta}{m} <\Psi_k^0|p_0^4|\Psi_k^0>
\label{pertmatrix}
\ee
For a central potential the unperturbed states are eigenstates of the angular momentum and thus we have $|\Psi_k^0>=|nlm>$ where $n$ counts the energy eigenstates and $l,m$ are the quantum numbers of angular momentum. 

Using the eigenvalue equation (\ref{eigval1}) in its unperturbed form it is straightforward to show that in each  subspace $(l,m)$ of given $n$, the matrix $<\Psi_k^0|p_0^4|\Psi_k^{\prime 0}>=<nlm|p_0^4|nl'm'>$ is diagonal and the first order correction to the spectrum is
\ba
\Delta E_{nl} &=& 4\beta m [(E_{nl}^0)^2-2E_{nl}^0\cdot <nlm|V(r)|nlm>\nn \\
&+& <nlm|V(r)^2|nlm>]
\label{denl}
\ea
where $m$ (the mass) should not be confused with the angular momentum quantum number in the bracket.
Assuming a power law central potential of the form
\be
V(r)\sim r^p
\label{vofr}
\ee
and using the virial theorem $<T>=\frac{p}{2}<V>$ we can write the first order correction (\ref{denl}) as
\be
\Delta E_{nl}=4\beta m \left[(E_{nl}^0)^2\frac{p-2}{p+2}+<nlm|V(r)^2|nlm>\right]
\label{denl1}
\ee 
Eq. (\ref{denl1}) is simple and general and can be used to derive the predicted shift in the spectrum in realistic systems like the hydrogen atom in the presence of a fundamental minimal position uncertainty \citep{Brau:1999uv-hydrogen,Benczik:2005bh-gup-hydrogen}. In the case of hydrogen atom with potential
\ba 
V(r)=\frac{\alpha_1}{r}
\label{hydropot}
\ea
where $\alpha_1$ is the fine structure constant, eq. (\ref{denl1}) leads to a shift of the energy spectrum of the form \cite{Brau:1999uv-hydrogen}
\be 
\Delta E_{nl}=\beta m^3 \alpha_1^4\frac{4n-3(l+1/2)}{n^4 (l+1/2)}
\label{hydrospectrum}
\ee
Clearly the shift of the energy eigenstates decreases rapidly for higher excited states. Thus the most sensitive state for measuring possible deviations from HUP is the ground state of the hydrogen atom.

\section{Maximum Length Quantum Mechanics}
\label{sec:Section 3}
\subsection{General Principles}
\begin{figure}[!t]
\centering
\vspace{0cm}\rotatebox{0}{\vspace{0cm}\hspace{0cm}
\resizebox{0.49\textwidth}{!}{\includegraphics{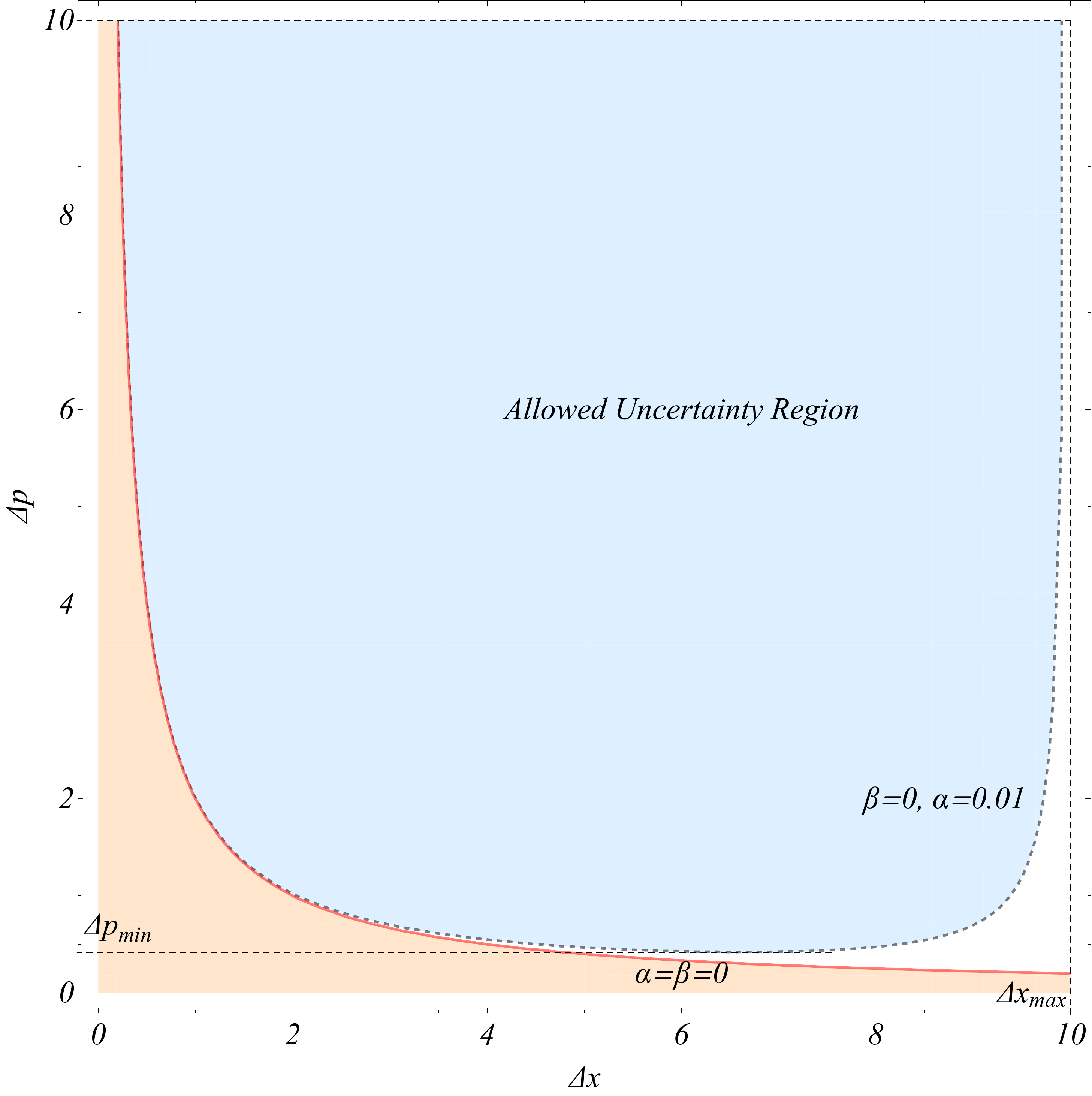}}}
\caption{The deformation of the HUP in accordance with eq. (\ref{gupexpmaxpos}) after rescaling to dimensionless form.}
\label{figgupminpmaxx}
\end{figure}

As discussed in the Introduction, a particularly general form of the GUP is expressed through eq. (\ref{gupexpmaxmommaxpos}) which includes explicit minima and maxima in both position and momentum. The allowed region of uncertainties in the context of this form of GUP is shown in Fig. \ref{figgupminmaxxp}. Motivated from the cosmological particle horizon or from possible nontrivial cosmic topology\citep{Luminet:2016bqv-cosmic-topology-constr} which provide a natural  maximum measurable length we now focus on the simple case of eq. (\ref{gupexpmaxmommaxpos}) with $\beta=0$ \ie without the presence of a minimum position uncertainty but with a maximum position uncertainty and a minimum momentum uncertainty in one space dimension (Fig. \ref{figgupminpmaxx}). We thus consider a commutation relation of the form
\be 
[x,p]=i\hbar \frac{1}{1 - \alpha x^2}\simeq i\hbar (1 + \alpha x^2)
\label{comrel1a}
\ee
where the last approximate equality is applicable under the condition $\alpha x^2\ll 1$. An operator representation that is compatible with the generalized commutation relation (\ref{comrel1a}) is 
\ba
p &=& \frac{1}{1 - \alpha x_0^2}p_0=(1+\alpha x_0^2 + \alpha^2 x_0^4 + ...)p_0
\label{reproperp} \\
x &=& x_0 
\label{reproperx}
\ea
It is straightforward to show that the commutation relation (\ref{comrel1a}) leads to a GUP of the form
\be 
\Delta x \Delta p \geq \frac{\hbar}{2} <\frac{1}{1-\alpha x^2}>\;\; \geq \; \frac{\hbar}{2}  \frac{1}{1- \alpha \Delta x^2}
\label{gupexpmaxpos}
\ee
which has similarities to eq. (\ref{gupexpmaxmom}) \cite{Pedram:2011gw-true-max-mom,Pedram:2012my-true-max-momii}. Clearly, the GUP (\ref{gupexpmaxpos}) indicates the existence of maximum position uncertainty 
\be 
l_{max}\equiv \Delta x_{max}= \frac{1}{\sqrt{\alpha}}
\label{maxposuncert}
\ee
It also has a minimum momentum uncertainty as can easily be verified by minimizing the uncertainty boundary equation
\be 
(1-\alpha \Delta x^2)\; \Delta x \Delta p - \frac{\hbar}{2} = 0
\label{uncertbound}
\ee
with minimum momentum uncertainty
\be 
\Delta p_{min}=\frac{3\sqrt{3}}{4} \hbar \sqrt{\alpha}
\label{minmomuncert}
\ee
which occurs when $\Delta x = 1/\sqrt{3\alpha}$ (see Fig. \ref{figgupminpmaxx} where $\alpha=0.01$).

\begin{figure*}[ht]
\centering
\begin{center}
$\begin{array}{@{\hspace{-0.10in}}c@{\hspace{0.0in}}c}
\multicolumn{1}{l}{\mbox{}} &
\multicolumn{1}{l}{\mbox{}} \\ [-0.2in]
\epsfxsize=3.3in
\epsffile{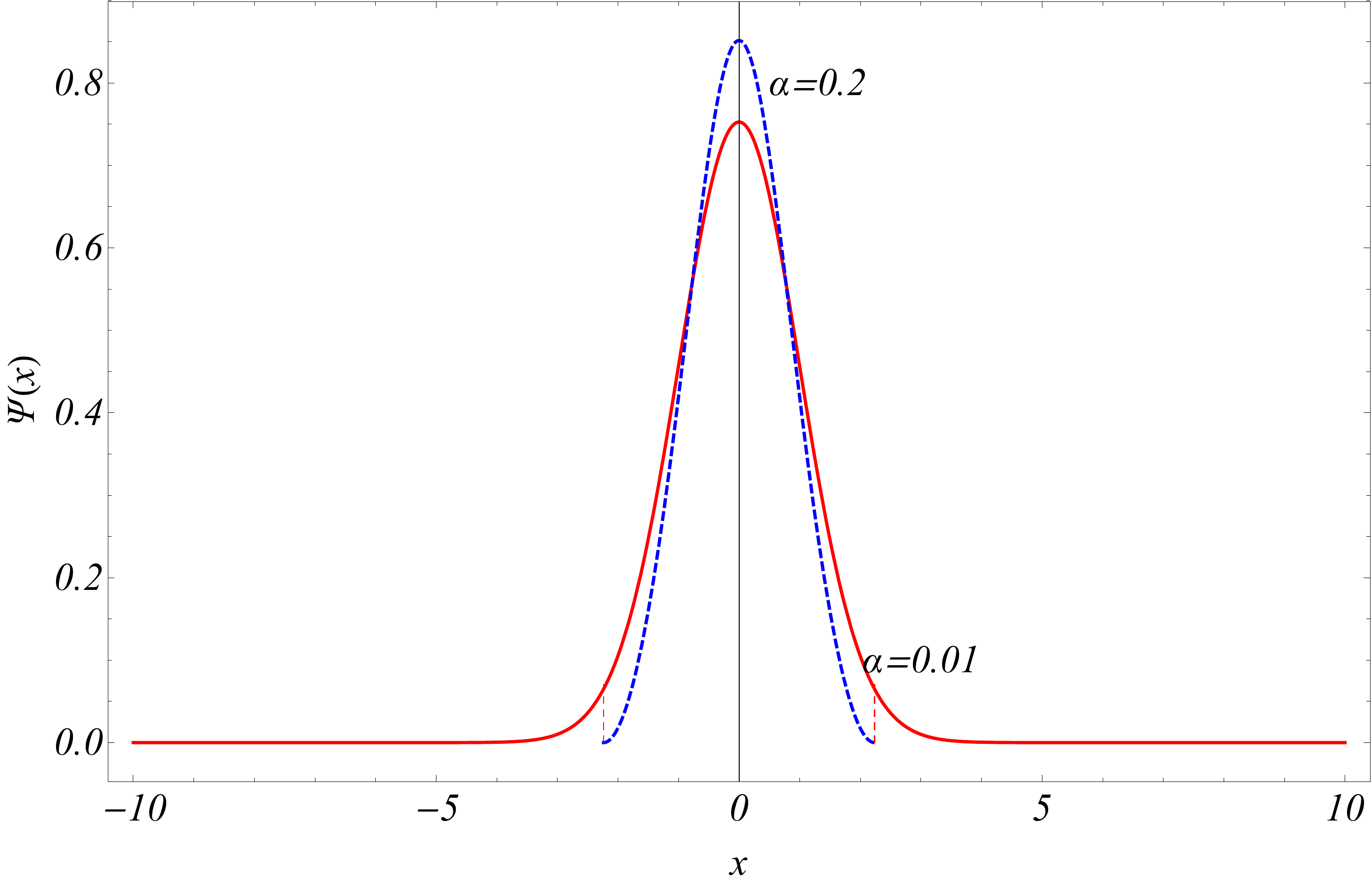} &
\epsfxsize=3.3in
\epsffile{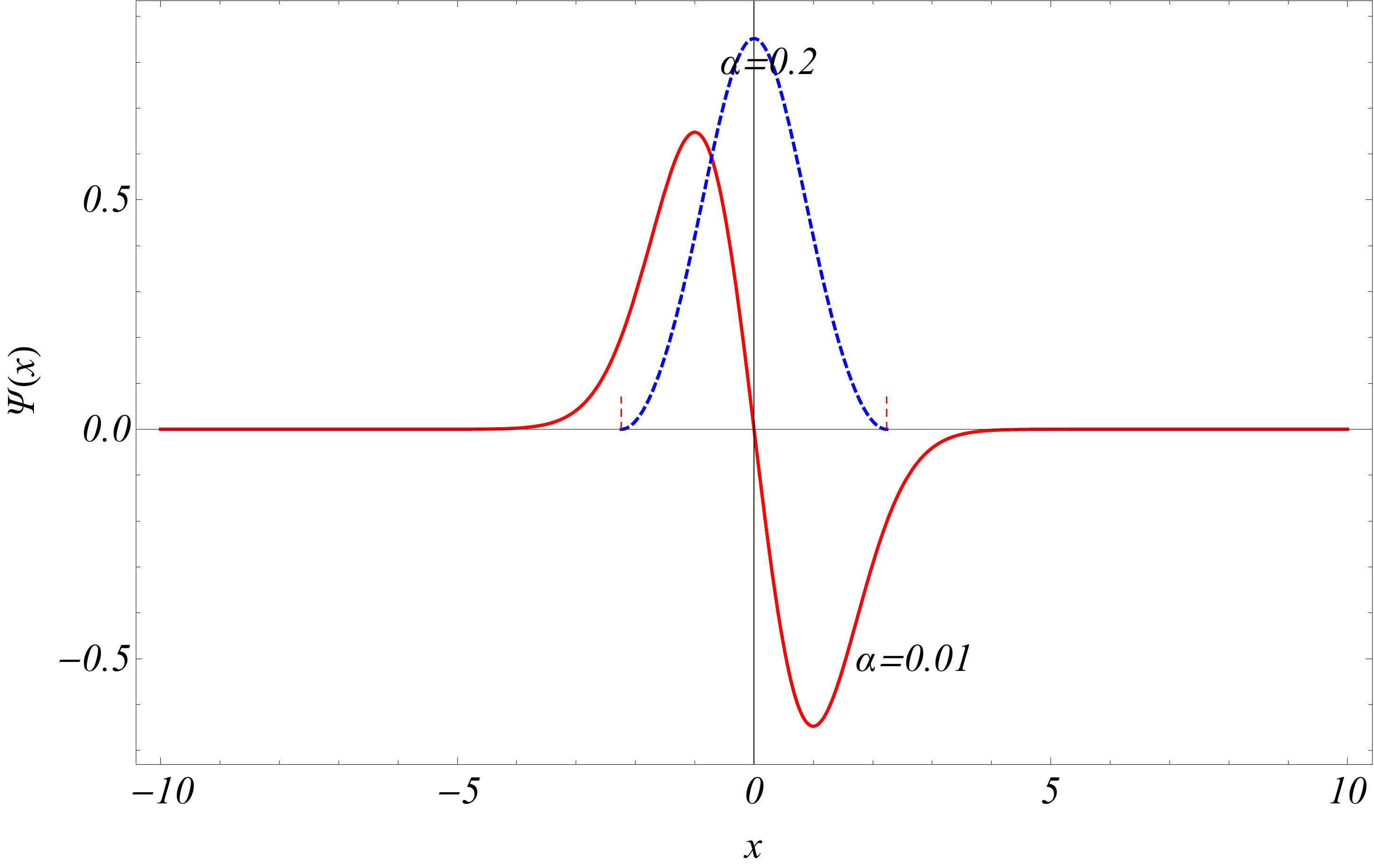} \\
\end{array}$
\end{center}
\vspace{0.0cm}
\caption{The ground state (left panel) and the first excited state (right panel) wavefunctions for $\alpha=0.01$ and for $\alpha =0.2$. The wavefunctions vanish at the IR cutoff $x=l_{max}=\frac{1}{\sqrt{\alpha}}$.}
\label{figwavefns}
\end{figure*}

\begin{figure*}[ht]
\centering
\begin{center}
$\begin{array}{@{\hspace{-0.10in}}c@{\hspace{0.0in}}c}
\multicolumn{1}{l}{\mbox{}} &
\multicolumn{1}{l}{\mbox{}} \\ [-0.2in]
\epsfxsize=3.3in
\epsffile{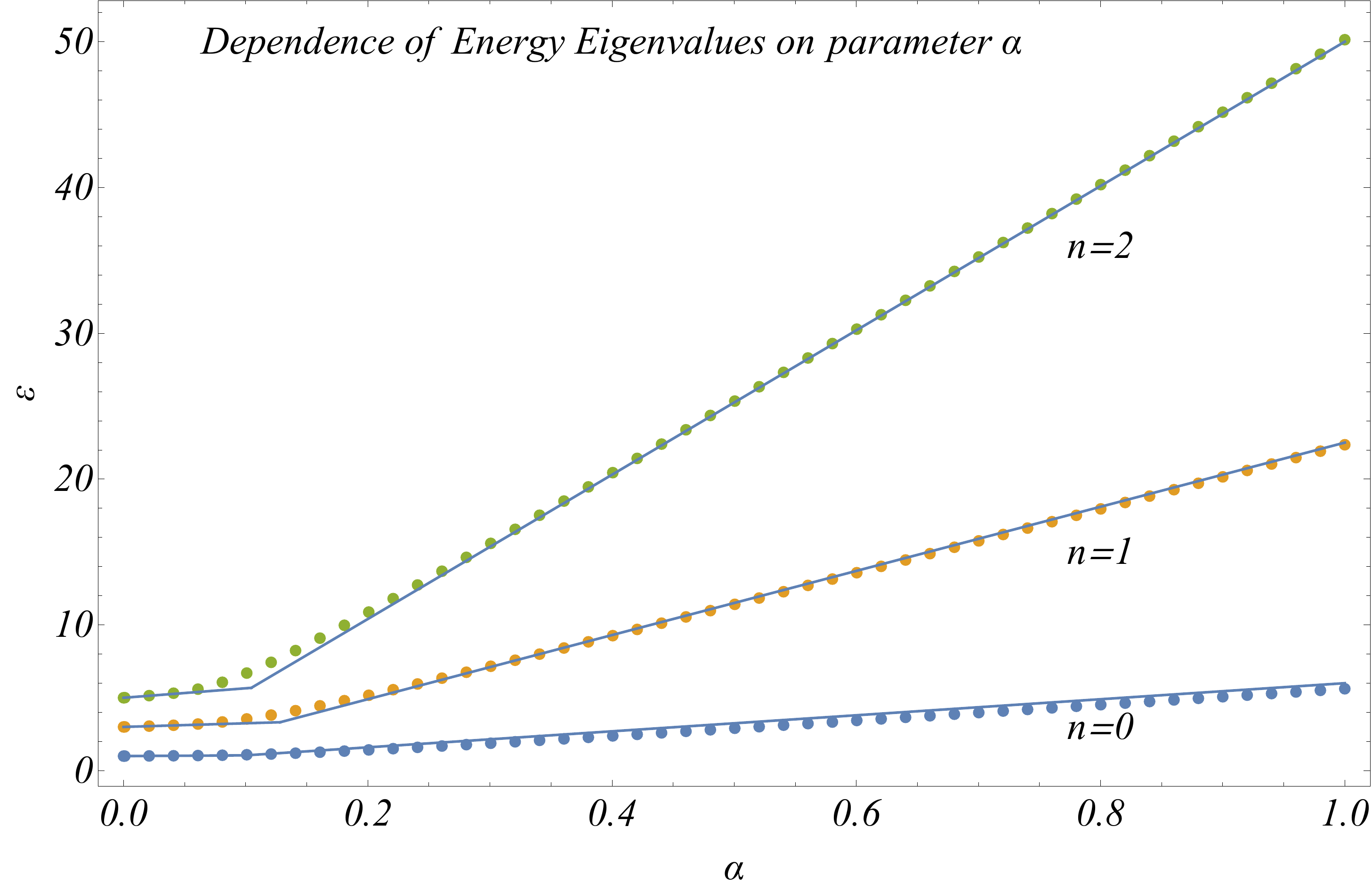} &
\epsfxsize=3.3in
\epsffile{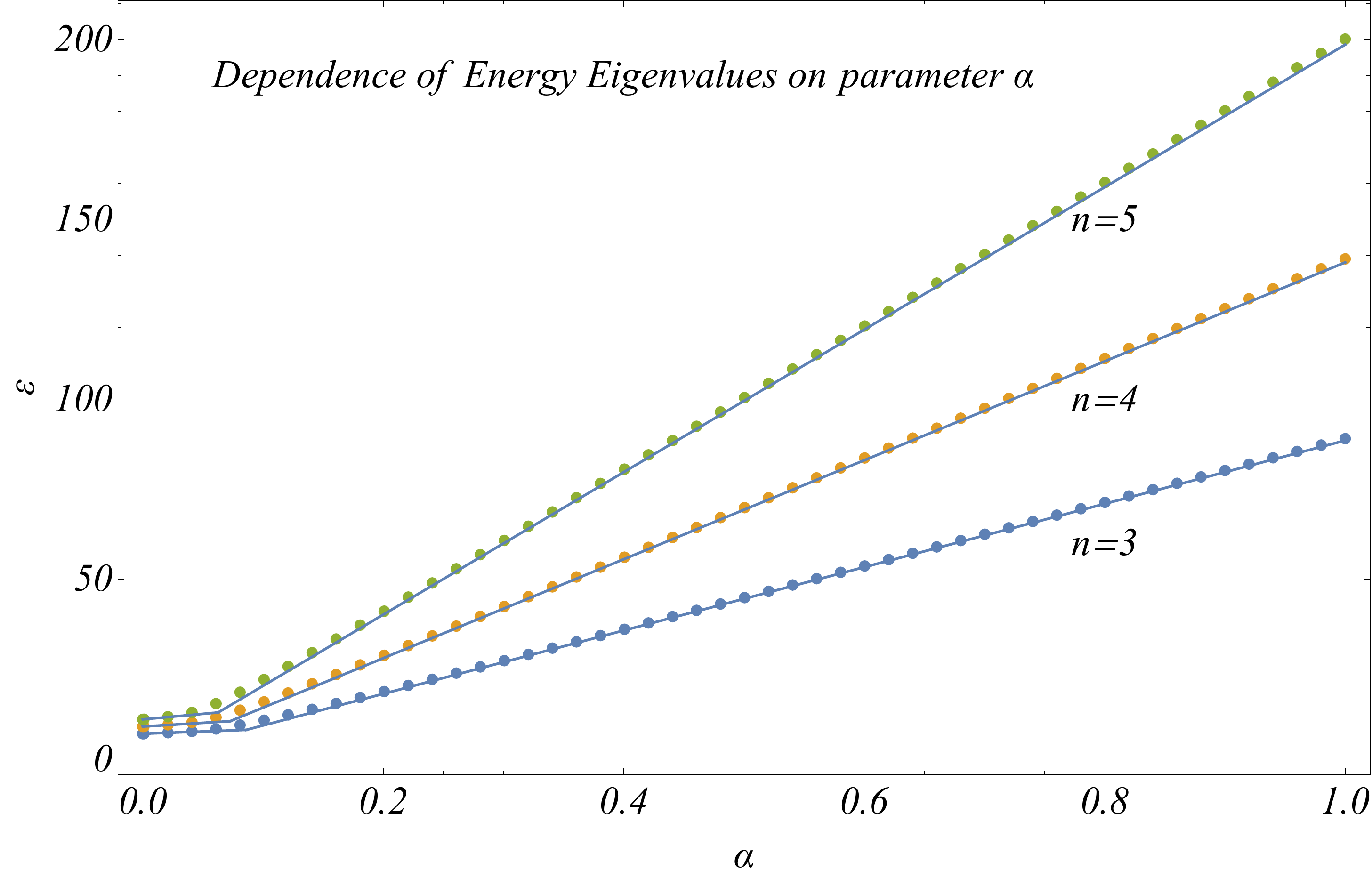} \\
\end{array}$
\end{center}
\vspace{0.0cm}
\caption{The dependence of the energy eigenvalues of the dimensionless parameter $\alpha$ for $n=0-6$. There is a linear dependence but the slope increases at a critical value $\alpha_{crit}(n)=\frac{4n+1}{11(n+1)^2-2n(n+1)-1}$. The slope for low $\alpha$ is well fit by the linear function $\mathcal{E}=2 n + 1 + (\frac{1}{2} + n (n + 1)) \alpha$ while the linear function for $\alpha \gg \alpha_{crit}$ is $\mathcal{E}=\frac{1}{2} + \frac{11}{2} (n + 1)^2  \alpha$. The thick dots correspond to the numerical results while the continous line is the analytical parametrization provided by (\ref{linsmall}), (\ref{linlarge}).}
\label{splens}
\end{figure*}

\begin{figure*}[ht]
\centering
\begin{center}
$\begin{array}{@{\hspace{-0.10in}}c@{\hspace{0.0in}}c}
\multicolumn{1}{l}{\mbox{}} &
\multicolumn{1}{l}{\mbox{}} \\ [-0.2in]
\epsfxsize=3.3in
\epsffile{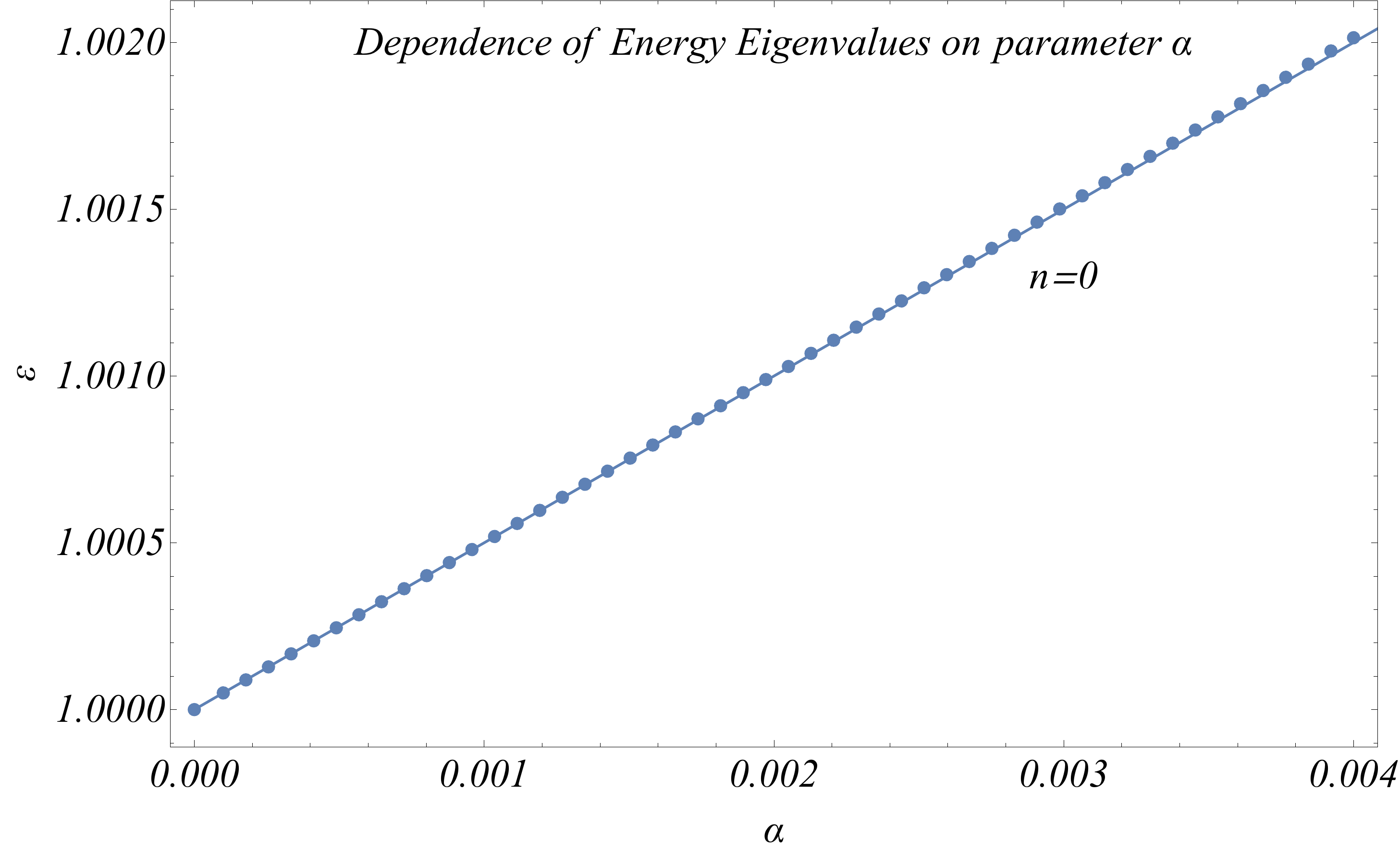} &
\epsfxsize=3.3in
\epsffile{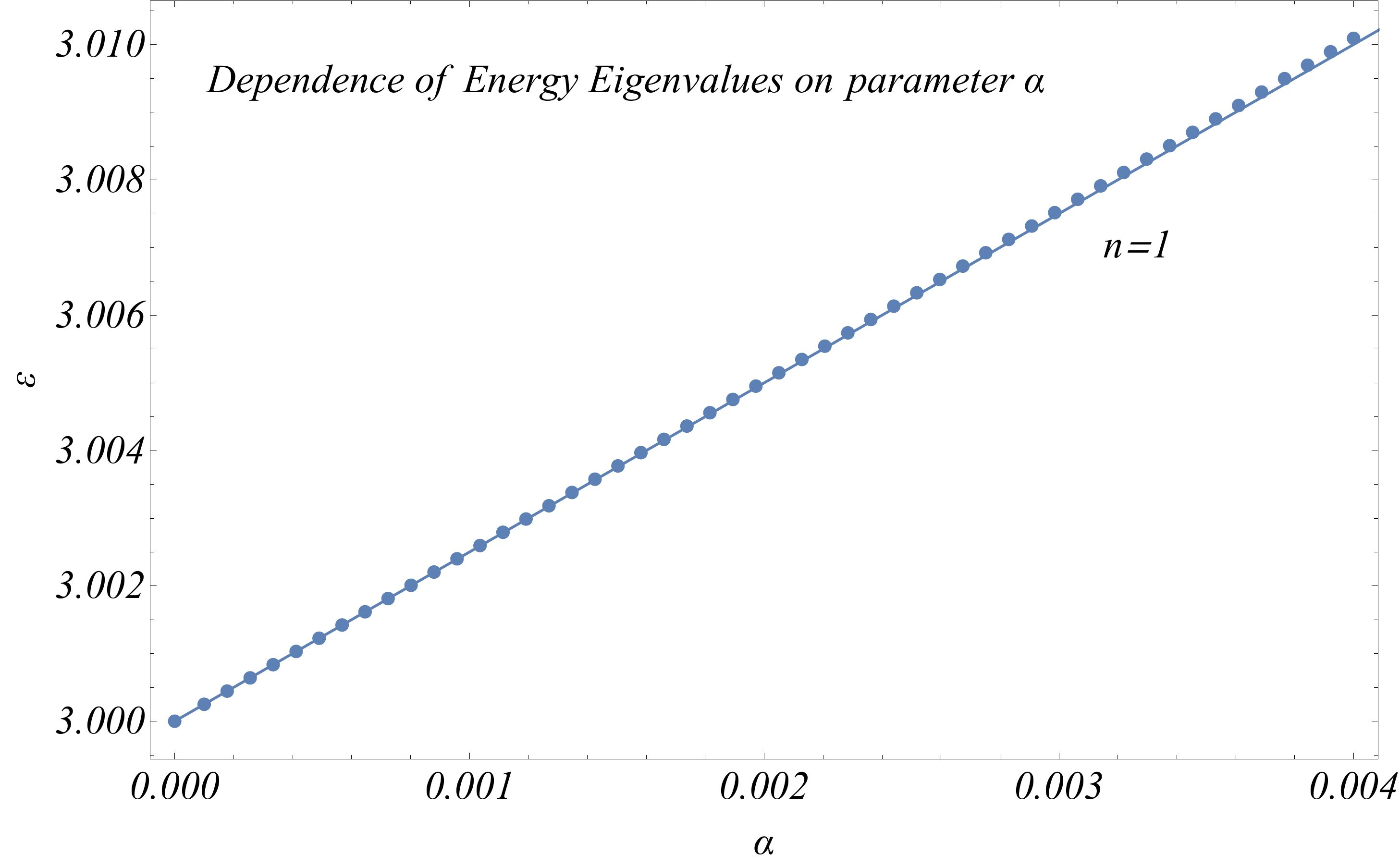} \\
\epsfxsize=3.3in
\epsffile{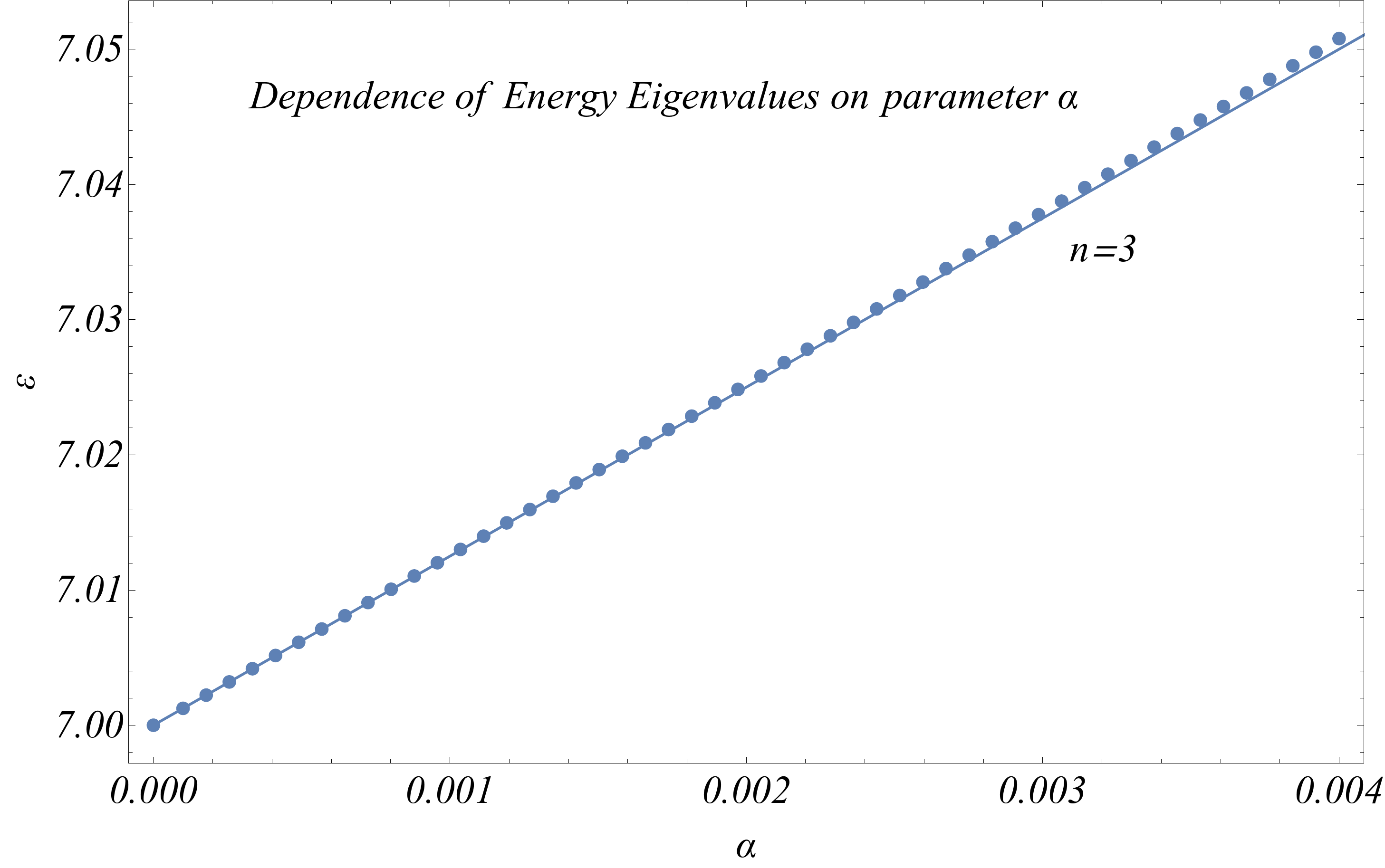} &
\epsfxsize=3.3in
\epsffile{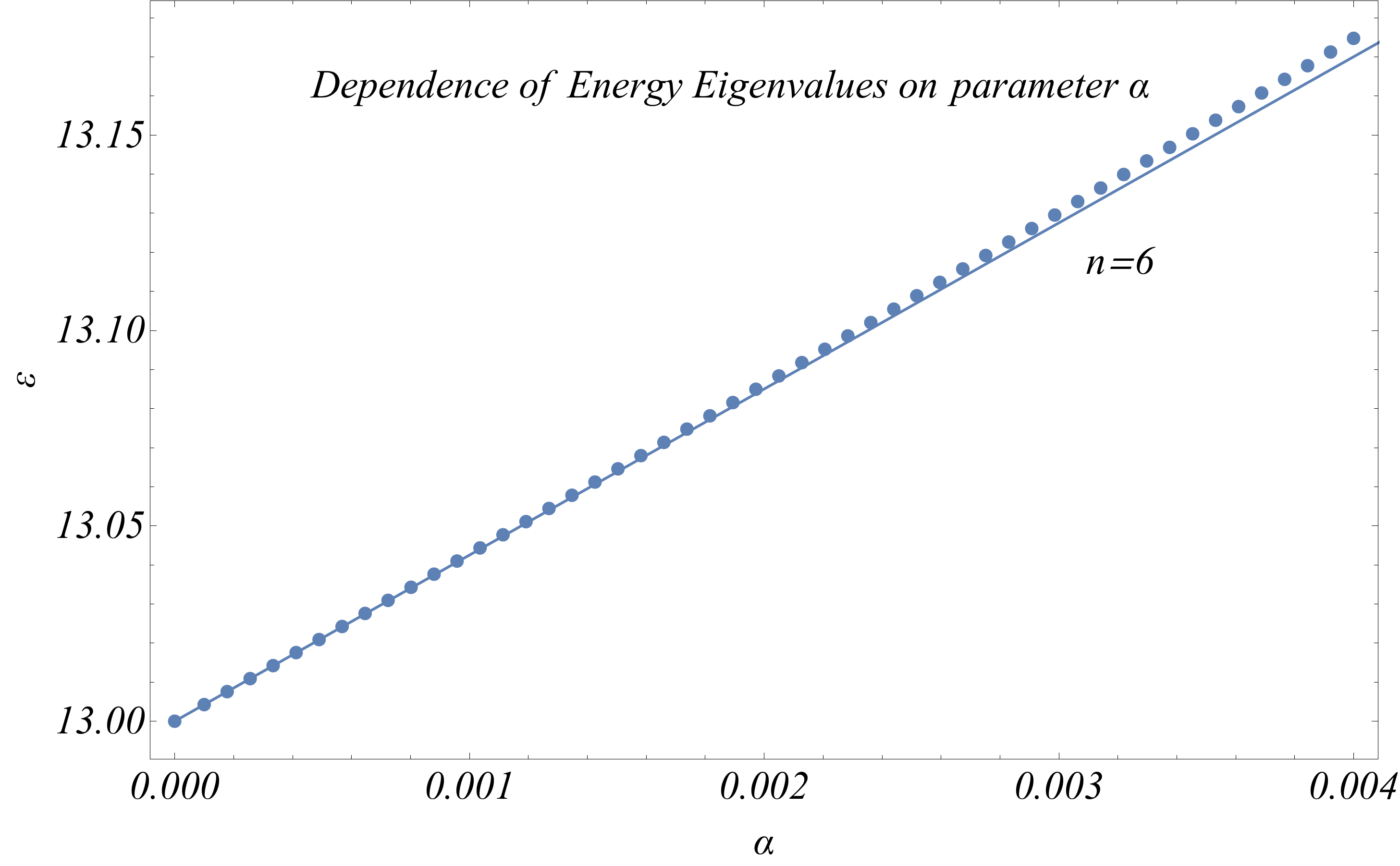}\\
\end{array}$
\end{center}
\vspace{0.0cm}
\caption{The dependence of the energy eigenvalues on the dimensionless parameter $\alpha$ for $n=0,1,3,6$ for low values of $\alpha$. Notice the linear behavior and the excellent fit with the analytical parametrization (\ref{linsmall}) (continous line). }
\label{splens1}
\end{figure*}

The GUP (\ref{gupexpmaxpos}) and the corresponding representation (\ref{reproperp}), (\ref{reproperx}) make specific predictions for deformation of the spectra of quantum mechanical systems that reduce to the HUP quantum mechanics in the limit $\alpha \rightarrow 0$. As discussed in the Introduction, a physically motivated maximum position uncertainty corresponds to the present day particle horizon given in eq. (\ref{parthort0}). Thus, by combining eqs. (\ref{maxposuncert}) and (\ref{parthort0}) we obtain a physically motivated value of the parameter $\alpha$ as
\be
\alpha = l_{max}^{-2} \simeq O(10^{-52})m^{-2}
\label{alphaphys}
\ee

Thus an important question that needs to be addressed is the following: 'What is the modification of the spectra of simple quantum systems induced in the context of the GUP (\ref{gupexpmaxpos}) and the corresponding representation (\ref{reproperp}), (\ref{reproperx}) for the physically motivated value of $\alpha$ given in eq. (\ref{alphaphys})?'

The answer of this question could also lead to the derivation of the specific signatures of the GUP (\ref{gupexpmaxpos}) in the spectra of physical systems and the potential of detectability of such signatures in present and future experiments. The presence of such signatures is a manifestation of the nonlocality of quantum mechanics which allows local systems to probe global properties of spacetime. 

Even though the parameter $\alpha$ is dimensionful, a relevant dimensionless parameter can be constructed for a given quantum system by rescaling $\alpha$ with the typical scale of the system. For example the typical microphysical length scale of a harmonic oscillator is 
\be 
x_{mp}=x_{osc}=\sqrt{\frac{\hbar}{m\omega}}\simeq O(10^{-12})m
\label{oscscale}
\ee
where in the last equality we have assumed the mass $m$ and the angular frequency $\omega$ corresponding to a typical diatomic molecule even though a charged particle in a homogeneous magnetic field  (Landau levels) could also be used as a physical system. Thus by combining eqs (\ref{alphaphys}),(\ref{oscscale}) we obtain a dimensionless version of $\alpha$ useful for the particular quantum system which may be written as
\be 
{\bar \alpha}\equiv \alpha x_{osc}^2 \simeq  O(10^{-77})
\label{alphabar}
\ee
which is extremely small. Even though the smallness of ${\bar \alpha}$ indicates that the corresponding deformation of the spectrum will turn out to be undetectable by current experiments it is still interesting to identify the predicted form of the spectral deformation and find the part of the spectrum that is mostly affected by this deformation. Thus  in the next subsection we focus on the derivation of this deformation in the simple harmonic oscillator in one spatial dimension.

\subsection{An Example: The Harmonic Oscillator}

The Hamiltonian of the one dimensional harmonic oscillator is of the form
\be 
H=\frac{p^2}{2m}+\frac{1}{2}m \omega^2 x^2
\label{hamiltosc}
\ee

Assuming the GUP (\ref{gupexpmaxpos}) and using the corresponding operator representation (\ref{reproperp}), (\ref{reproperx}) the Hamiltonian takes the form
\be 
H=\frac{1}{2m}\frac{1}{1-\alpha x_0^2} p_0 \frac{1}{1-\alpha x_0^2} p_0 +\frac{1}{2}m \omega^2 x_0^2
\label{hamiltosc}
\ee
In position space the undeformed momentum operator $p_0$ takes the form
\be 
p_0=-i\hbar \frac{d}{dx_0}=-i\hbar \frac{d}{dx}
\label{momoperposspa}
\ee
Using eqs (\ref{hamiltosc}) and (\ref{momoperposspa}) it is straightforward to show that the Schrodinger equation $H\Psi(x)=E\Psi(x)$ in position space takes the generalized form
\be 
\frac{d^2 \Psi}{dx^2}+\frac{2\alpha x}{1-\alpha x^2}\frac{d\Psi}{dx}+(1-\alpha x^2)^2 \left(\mathcal{E}-\eta x^2\right)\Psi=0
\label{schreq1}
\ee
where $\mathcal{E}\equiv \frac{2mE}{\hbar^2}$, $\eta\equiv \frac{m\omega}{\hbar}$ and $x^2\in [0,\frac{1}{a}]$. This equation corresponding to an IR cutoff,  is formally the same as the corresponding equation obtained when an explicit UV cutoff is imposed (explicit maximum momentum uncertainty)\citep{Pedram:2011gw-true-max-mom} and may be studied using similar approximate analytical methods. However, here we choose the use of numerical methods as they lead to better description of the global behaviour of the solutions. 

Clearly there are two scales in eq. (\ref{schreq1}): the microphysical system scale $x_{osc}^2\equiv \frac{1}{\eta}$ and the fundamental GUP scale $l_{max}^2=\frac{1}{\alpha}$. We now define the dimensionless quantities ${\bar x}\equiv x \sqrt{\eta}$, ${\bar \alpha}\equiv \frac{\alpha}{\eta}$ and 
\be 
\mathcal{{\bar E}}\equiv \frac{\mathcal{E}}{\eta}=\frac{E}{\frac{1}{2}\hbar \omega}
\label{edimless}
\ee 
In the absence of maximal position uncertainty ($\alpha=0$) $\mathcal{{\bar E}}(n)=2 n+1$. 
Using these quantities, the Generalized Schrodinger Equation (GSE) of (\ref{schreq1}) may be written in dimensionless form as
\be 
\frac{d^2 \Psi}{d\bar x^2}+\frac{2\bar \alpha \bar x}{1-\bar \alpha \bar x^2}\frac{d\Psi}{d\bar x}+(1-\bar \alpha \bar x^2)^2 \left(\mathcal{\bar E}- \bar x^2\right)\Psi=0
\label{schreq2}
\ee
The rescaled GSE (\ref{schreq2}) involves a single dimensionless parameter $\bar \alpha$.  In what follows we omit the bar for simplicity. It is straightforward to solve the GSE (\ref{schreq2}) numerically using Mathematica \cite{math10m} under the following boundary conditions:
\begin{itemize}
\item 
The wavefunctions should vanish at the maximal position $l_{max}=\frac{1}{\sqrt{\alpha}}$ ($\Psi(l_{max})=0$) as indicated by the divergence of the effective potential in eq (\ref{schreq2}).
\item 
The wavefunction should have definite parity ($\Psi(x)=\pm \Psi(-x)$)
\item 
The wavefunction should be properly normalized ($\int_{-l_{max}}^{+l_{max}} (1-\alpha x^2) \vert \Psi(x) \vert^2 dx =1$). The term $(1-\alpha x^2)$ in the scalar product definition used in the normalization is needed in order to retain symmetricity of the momentum operator defined in eq. (\ref{reproperp}) \cite{Kempf:1994su-basic,Pedram:2011gw-true-max-mom}.
\end{itemize}
These conditions are sufficient to lead to both the energy spectrum and the wavefunctions for any value of $\alpha$. In Fig. \ref{figwavefns} we show the ground state and the first excited state normalized wavefunctions for $\alpha=0.01$ and for $\alpha =0.2$. As imposed by the boundary conditions, the wavefunctions vanish at $x=l_{max}=\frac{1}{\sqrt{\alpha}}$. Notice the confinement of the wavefunction for larger values of $\alpha$ which leads to increased energy eigenvalues.

The energy spectrum $\mathcal{ E}(n,\alpha)$ may also be evaluated numerically using the above boundary conditions and eq. (\ref{schreq2}). The energy eigenvalues as a function of the dimensionless parameter $\alpha$ for $n=0-6$ are shown in Fig. \ref{splens} (thick dots). Clearly the dependence of the eigenvalues on $\alpha$ is linear for both small and {\it large} values of $\alpha$. The slope of the linear dependence however changes at a critical value $\alpha_{crit}$ that depends on the value of the quantum number $n$. It is straightforward to show using Mathematica \cite{math10m} that the linear dependence of the energy eigenfunctions on $\alpha$ may be very well approximated by the following parametrization
\ba 
\mathcal{E}(n,\alpha) &=& 2 n + 1 + (\frac{1}{2} + n (n + 1)) \alpha, \hspace{0.1cm} \alpha <\alpha_{crit} 
\label{linsmall}
\\
\mathcal{E}(n,\alpha) &=&\frac{1}{2} + \frac{11}{2} (n + 1)^2  \alpha,  \hspace{1.3cm} \alpha > \alpha_{crit}
\label{linlarge}
\ea
where 
\be 
\alpha_{crit}(n)=\frac{4n+1}{11(n+1)^2-2n(n+1)-1}
\label{acrit}
\ee
The quality of fit of the parametrization (\ref{linsmall}), (\ref{linlarge}) to the numerically obtained energy spectrum is demonstrated in Fig. \ref{splens} where we superpose the numerically obtained eigenvalues (thick dots) for various values of $n$ with the corresponding linear relations (\ref{linsmall}), (\ref{linlarge}) (continous lines). The linear relation for small $\alpha$ is particularly interesting in view of the physical arguments leading to eq. (\ref{alphabar}). The energy eigenvalues in this range of $\alpha\ll \alpha_{crit}$ is shown in Fig. \ref{splens1} along with the corresponding fits of eq (\ref{linsmall}) which clearly provide an excellent fit to the numerically obtained eigenvalues $\mathcal{E}(n,\alpha)$.

\begin{figure}[!t]
\centering
\vspace{0cm}\rotatebox{0}{\vspace{0cm}\hspace{0cm}
\resizebox{0.49\textwidth}{!}{\includegraphics{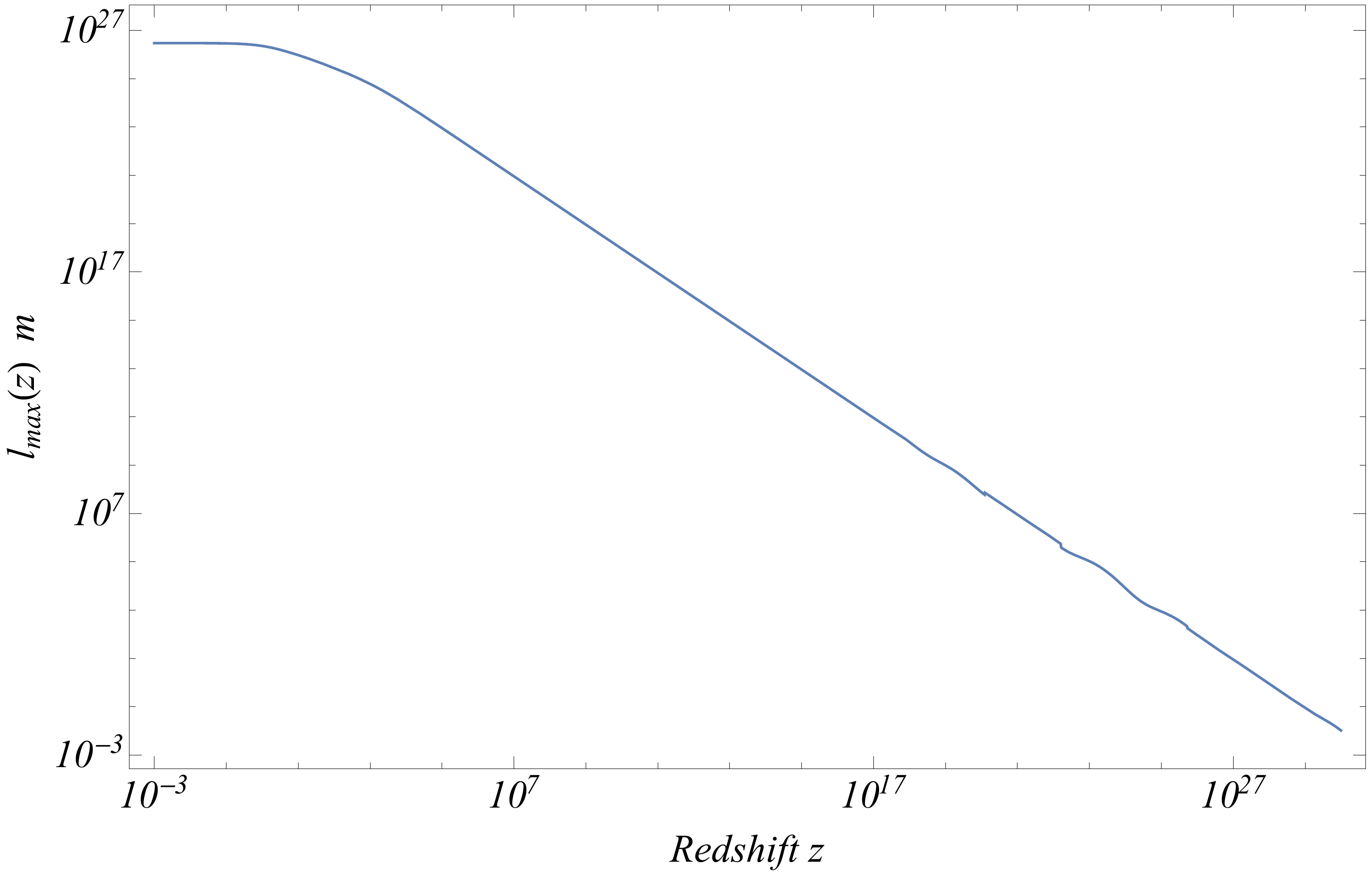}}}
\caption{The comoving particle horizon or maximum position uncertainty in a \lcdm universe vs redshift. At high redshifts the maximum position uncertainty becomes microphysical.}
\label{figxmaxz}
\end{figure}

\subsection{Time dependence of maximum position uncertainty}
If the maximum position uncertainty $l_{max}$ (and therefore $\alpha$) is assumed to be determined by the comoving particle horizon then it should be a time dependent quantity on cosmological timescales. This time dependence is expressed in a cosmological setup as a scale factor $a$ or redshift $z$ dependence. We thus have
\be 
l_{max}(z)=c \int_0^z \frac{dz}{H(z)}
\label{xmaxz}
\ee
where $H(z)$ is the redshift dependent Hubble expansion rate which in \lcdm takes the form
\be
H(z)=H_0  \sqrt{\Omega_{0m} (1+z)^3 + \Omega_{0r} (1+z)^4 +\Omega_\Lambda}
\label{hz}
\ee
and the Hubble radius is
\be
\frac{c}{H_0}\simeq  9\times 10^{25} h \;\; meters
\label{hubrad}
\ee
while $h$ is the Hubble parameter in units of $100 km/(sec\cdot Mpc)$. Using eqs (\ref{xmaxz})-(\ref{hubrad}) it is straightforward to obtain the maximum position uncertainty $l_{max}(z)$ vs redshift (in meters) assuming a \lcdm universe with $\Omega_{0m}=0.3$, $\Omega_{0r}=10^{-4}$. Such a log-plot of $l_{max}(z)$ is shown in Fig. \ref{figxmaxz}. Clearly at high redshift the maximum position uncertainty becomes microphysical and may produce signatures in the quantum fluctuations produced during inflation leading to structure formation. In particular a generalized commutation relation of the form (\ref{comrel1a}) is expected to also modify the commutation relation between creation and annihilation operators of the harmonic oscillator (generalized bosonic Heisenberg algebra \cite{Kempf:1993bq,Kempf:1994qp}) leading also to quantum field theoretical effects. These effects are unobservable at the present universe but at the early universe they may lead to observable deviations from the scale invariant primordial power spectrum generated during inflation. The investigation of these effects is beyond the scope of the present analysis.

\section{Conclusion-Early Universe Signatures}
\label{sec:Section 5}

We have demonstrated that the existence of a maximal position uncertainty leads to nontrivial modifications of the properties of local quantum systems due to the nonlocality that is inherent in quantum mechanics. The existence of such a maximal postion uncertainty is generic in a cosmological setup due to the presence of particle horizons. 

If the maximal position uncertainty is as large as the present particle horizon of the univerce then its effects on local microphysical quantum systems like the harmonic oscillator exist but they are not large enough to be observable. However, in the early universe when the comoving particle horizon is much smaller than its present size, the effects of a maximum position uncertainty may be important thus leaving a signature on the shape of the primordial power spectrum of quantum cosmological fluctuations generated during inflation.

These results are generic, model independent and are generated simply by demanding consistency of quantum mechanics with the description of the universe in the context of Big Bang cosmology. Their increased importance in the physical processes of the early universe make them particularly interesting and raises the possibility of the existence of observational signatures in cosmological data. This possibility leads to a wide range of possible extensions of the present work. These extensions include the following
\begin{itemize}
\item
{\bf 3D Systems:} Investigate the spectrum modifications induced by the presence of maximal position uncertainty in three dimensional quantum systems like the hydrogen atom.
\item
{\bf Field Theoretical Effects:} Study the predicted effects of maximal position uncertainty in the context of field theory\cite{Kempf:1994qp} and the predicted modifications induced in scattering amplitudes and path integral\cite{Pramanik:2014mma} formalism.
\item
{\bf Early Universe Signatures:} The effects of a maximum position uncertainty due to particle horizon on non-trivial cosmic topology are expected to be amplified in the Early Universe and lead to observable effects in the context of Nucleosynthesis, Primordial fluctuations spectrum, effects on thermal equilibrium etc.
\item 
{\bf Simultaneous presence of Maximal and Minimal position uncertainty:} As stated  in the Introduction, quantum gravitational considerations imply the existence of a minimal position uncertainty. The behaviour of quantum systems in the simultaneous presence of  Maximal and Minimal position uncertainties (eq. (\ref{gupexpmaxmommaxpos})) is also an interesting extension of the present analysis.
\end{itemize}

{\bf Numerical Analysis:} The Mathematica file that led to the production of the figures may be downloaded from \href{http://leandros.physics.uoi.gr/maxlenqm/}{here}.

\section*{Acknowledgements}
I thank Elias Vagenas for interesting discussions.

\raggedleft
\bibliography{bibliography}

\end{document}